%% file: 00-main.tex
\documentclass[sigconf]{acmart}

\usepackage{booktabs} % For formal tables
\usepackage{textgreek}
\usepackage{multirow}

\setcopyright{acmlicensed}
\acmDOI{10.1145/3219104.3219131}

\acmISBN{978-1-4503-6446-1/18/07}

\acmConference[PEARC'18]{ACM Practice and Experience in Advanced Research Computing}{July 22--26,2018}{Pittsburgh, PA, USA}

\acmYear{2018}
\copyrightyear{2018}

\acmPrice{15.00}

\begin{document}
\title{Virtualizing the Stampede2 Supercomputer with Applications to HPC in the Cloud}

\author{W. Cyrus Proctor}
\affiliation{%
  \institution{Texas Advanced Computing Center}
  \streetaddress{10100 Burnet Road}
  \city{Austin}
  \state{Texas}
  \postcode{78758}
}
\email{cproctor@tacc.utexas.edu}

\author{Mike Packard}
\affiliation{%
  \institution{Texas Advanced Computing Center}
  \streetaddress{10100 Burnet Road}
  \city{Austin}
  \state{Texas}
  \postcode{78758}
}
\email{mpackard@tacc.utexas.edu}

\author{Anagha Jamthe}
\affiliation{%
  \institution{Texas Advanced Computing Center}
  \streetaddress{10100 Burnet Road}
  \city{Austin}
  \state{Texas}
  \postcode{78758}
}
\email{ajamthe@tacc.utexas.edu}

\author{Richard Cardone}
\affiliation{%
  \institution{Texas Advanced Computing Center}
  \streetaddress{10100 Burnet Road}
  \city{Austin}
  \state{Texas}
  \postcode{78758}
}
\email{rcardone@tacc.utexas.edu}

\author{Joseph Stubbs}
\affiliation{%
  \institution{Texas Advanced Computing Center}
  \streetaddress{10100 Burnet Road}
  \city{Austin}
  \state{Texas}
  \postcode{78758}
}
\email{jstubbs@tacc.utexas.edu}

\renewcommand{\shortauthors}{Proctor et al.}

\begin{abstract}
Methods developed at the Texas Advanced Computing Center (TACC) are described and demonstrated for automating the construction of an elastic, virtual cluster emulating the Stampede2 high performance computing (HPC) system. The cluster can be built and/or scaled in a matter of minutes on the Jetstream self-service cloud system and shares many properties of the original Stampede2, including: i) common identity management, ii) access to the same file systems, iii) equivalent software application stack and module system, iv) similar job scheduling interface via Slurm. 

We measure time-to-solution for a number of common scientific applications on our virtual cluster against equivalent runs on Stampede2 and develop an application profile where performance is similar or otherwise acceptable. For such applications, the virtual cluster provides an effective form of ``cloud bursting'' with the potential to significantly improve overall turnaround time, particularly when Stampede2 is experiencing long queue wait times. In addition, the virtual cluster can be used for test and debug without directly impacting Stampede2. We conclude with a discussion of how science gateways can leverage the TACC Jobs API web service to incorporate this cloud bursting technique transparently to the end user.
\end{abstract}




\date{26 March 2018}
\maketitle

\input{01-intro}

\input{02-implementation}
\input{03-results}
\input{04-conclusions}

%ACKNOWLEDGMENTS are optional
\section{Acknowledgments}
We thank Harika Gurram for reconfiguring her production Agave definitions to allow our experiments to run.  

\bibliographystyle{ACM-Reference-Format}
\bibliography{05-references}

\end{document}

%% file: 01-intro.tex
\section{Introduction}
\label{introduction}

As the demand for HPC continues to grow, centers must cater to a deep range of researchers bringing forward more numerous and challenging workflows. The ability to automatically and intelligently respond when computing needs outstrip a current system's supply provide a pathway towards reducing overall time-to-solution. With mutiple, on-premises production systems available at TACC, we are investigating the potential for automatically offloading work from one system to another in ways that are largely transparent to the user.

In particular, this work catalogues and demonstrates a set of necessary components for the creation of a virtualized HPC execution environment on the Jetstream cloud system. Ultimately, this setup emulates the environment found on the Stampede2 HPC system allowing for the potential of on-demand overflow capacity with effectively no user-level adaption required.

\subsection{Background}
Although the concept of virtualization has been around for decades \cite{reuther2012}, HPC virtualization for industrial and scientific research is gaining significant attention recently. Huang et al. \cite{huang2006} in a case study, describe the benefits of HPC virtualization. Virtualization can effectively address the issues of system management and allow scalability by cloud bursting. System administrators can easily configure the required runtime setup and spin up VMs to run various applications. Hardware upgrades and failures on large scale systems can be handled gracefully, using a checkpoint/restart mechanism. Also, operating systems and runtime environments can be largely customized to gain optimal performance, beneficial for HPC applications.

Walters et al. \cite{walters2008} discuss three major categories of virtualization including full virtualization, paravirtualization ,and operating system level virtualization. Their work evaluates the suitability of these techniques for industry standard VMs (VMWare Server, Xen, and OpenVZ). Their test setup focuses on evaluating performance of various VMs running HPC applications for file reads/writes, network utilization and other scientific benchmarks: symmetric multiprocessor performance and MPI scalability. Although the performance of the base system is not exactly matched by any of the VMs, the authors point out potential areas where the systems perform well and scope of improvement.

Guo et al. \cite{GuoSSWS14} describe a predictive cost model that optimizes cloud bursting to a remote cluster.  Applications that run in one or more VMs in an enterprise's private cloud can be moved to a public cloud when more resources are required.  Their predictive model uses integer linear programming and heuristics to determine when and which VMs to move.  A important consideration in their strategy is how to manage the movement of large VMs and other data across low bandwidth networks.  The problem addressed in this paper is how to achieve reasonable time-to-solution for HPC batch jobs when the overflow cloud resources are colocated with the HPC system and data movement is not a primary consideration.  

Multiple studies analyze the feasibility of running computationally intensive scientific applications on commodity cloud \cite{he2010,rehr2010}, such as Amazon EC2 \cite{ec2amazon2006}. However, not all public clouds are capable of running HPC applications due to network latency or lack of fast interconnection between the virtual machines. Even so, some of the previous experiments have obtained satisfactory results when HPC applications were run on Amazon EC2 \cite{he2010}. Unlike these studies, in this paper, we chose to run HPC applications by virtualizing the Stampede2 system \cite{stanzione2017} which has the proven capability to successfully run hundreds of computing intensive scientific applications.

\subsection{Contributions}

This paper outlines a method for constructing a virtual HPC cluster in an elastic cloud environment that uses disparate hardware and system management techniques when compared with the original HPC system. The method addresses issues common to any HPC virtualization effort and explains the particular choices tailored for TACC's Stampede2 system. Initial experiments comparing HPC and cloud time-to-solution for some common scientific applications provide proof-of-concept that cloud bursting HPC workloads can be practical and effective. Moreover, this work demonstrates of how a job scheduler or middleware such as Agave can seamlessly be used to tie HPC and cloud environments.

The rest of this paper is organized as follows:  Section \ref{implementation} provides implementation details for our setup, including the system configuration, virtualization techniques, and platform details. Section \ref{results} describes the experimental setup and time-to-solution results for the HPC and cloud systems. In Section \ref{conclusions}, we conclude this paper and discuss scope for future work.

%% file: 02-implementation.tex
\section{Implementation}
\label{implementation}

To provide a virtual extension to existing HPC infrastructure, it is necessary to identify core software and hardware components that allow for a specific design level of interoperability. The tighter the integration, the more seamless the experience of the intended audience. This section introduces key components utilized in and tailored for both the Stampede2 HPC system and the Jetstream self-service cloud system. Additional components crafted specifically for this work that ultimately provide a consistent software experience across systems are also discussed.

\subsection{Basic System Configuration}
\label{sysconfig}

System management at large scale has two main thrusts: provisioning and change management. Together, these two applied concepts are responsible for maintaining the state of the cluster. Generally, the provisioning step selects resources from a pool of network servers to load the appropriate software (including operating system, device drivers, middleware, and basic applications), customize unique network information, and associate storage resources. Once a server or VM is provisioned, a change management application functions to provide incremental software updates beyond the initial state of the machine.

At its heart, Stampede2 uses Cobbler and LosF \cite{mclay2011,cobbler,losf} for its provisioning and change management systems while Jetstream utilizes a combination of OpenStack and Ansible \cite{openstack,ansible}. The divergence in system administration approches between these systems rule out a more natural extension of simply adding compute resources to an existing sever pool. 

At a hardware level, Stampede2 and Jetstream have little in common. For the purposes of this discussion, Stampede2 consists of three different node classes: master, compute, and login. In the compute class, there are 4200 single socket Intel Xeon Phi 7250 (KNL) nodes with 16GB of MCDRAM plus 96GB DDR4 RAM.  In addition, there are 1736 2-socket Intel Xeon Platinum 8160 (SKX) nodes with 192GB of DDR4 RAM.  These nodes are connected via a 100Gb/s Intel Omni-Path (OPA) network in a 7:5 oversubscribed fat tree topology. The front-facing login node class consists of 8 2-socket Intel Xeon Gold 6132 (SKX) nodes with 96GB DDR4 RAM each connected into the OPA network and serving the outside world via two network-bonded 10 gigabit Ethernet connections. The master node is a single 2-socket Intel Xeon CPU E5-2680 v4 (BDW) node with 96GB of DDR4 RAM also connected to the OPA network but isolated from the outside. Connecting all nodes, three main parallel Lustre file systems referred to as \texttt{/home}, \texttt{/work}, and \texttt{/scratch} can provide single-job aggregate performance write I/O of 300GB/s across 512 nodes with a total capacity of more than 30PB.

Jetstream's hardware at TACC consists of 320 Intel Xeon E5-2680 v3 (HSW) compute nodes with 128GB DDR4 RAM that are connected at 10Gb/s to a 40Gb/s Ethernet backbone, with a shared external connection of 120Gb/s. Each node contains 2TB of local storage as well as access to a 960TB network attached global Ceph storage system. An additional factor unique to Jetstream is that multiple virtual machines may be colocated on each compute node necessitating the need for sharing of resources including the CPU, memory, and network interface card.

At a basic software level, both Jetstream and Stampede2 are capable of running the same Linux CentOS 7.4.1708 distribution. Stampede2 is heavily customized with over 250 staff-supported software packages in addition to several hundred more that are community-provided in the form of RPMs. These RPMs are installed via LosF in addition to the standard OS distribution RPMs provided via Cobbler. Jetstream VMs, on the other hand, start from community developed and maintained collections of RPMs that are tailored to specific functions that a user may need. Generally, the sets of RPMs tend to be close to the original OS distribution RPM set with additions made from online community repositories.

\subsection{Virtualization Ingredients}

For a seamless virtual extension, Jetstream needed to share or emulate a few common features of Stampede2. Some of these include the shell start-up environment, file systems, batch scheduler, identity management, and internode communication pathways. Depending on how similar these features are created on the cloud extension system dictate the difference between packaging up an autonomous unit of computing work to be run on a completely generic cloud to that same unit of work being run on a system that is practically indistinguishable from the original HPC system.

Identity management for both systems can be synchronized through the use of TACC's LDAP directory service to ensure that users exist with the same name and group structures on both machines. This is facilitated through administrative queries to a set of servers that maintain up-to-date records site-wide.

The ability to share user and group information between systems allowed for a straightforward mechanism to present Stampede2 file systems in a logical and secure manner on Jetstream. For the purposes of this demonstration, the \texttt{/home}, \texttt{/work}, and \texttt{/scratch} Lustre file systems were re-exported via NFS as well as the locations of the majority of staff-supported software in the local \texttt{/opt} directory. Jetstream VMs were then free to mount these file systems upon start-up to provide a functional, if not as performant, approach to interacting natively with files and applications on Stampede2.

Files that customize and present the user with a unique shell environment upon login as well as other crucial system configuration files also need to be transferred to the cloud extension system. These files, which constitute the TACC user environment, have been honed over several generations of compute systems to provide a scalable and flexible platform to meet a wide range of computing needs. A minimum core of RPMs were identified and outfitted for use on a generic cloud platform and served out from a custom-built Yum repository. This configuration allows for any node or VM with a connection to the outside world to import this TACC repository and its associated signed GPG key and install these RPMs as part of a change management step.

The other key set of RPMs that were presented from the TACC repository included the basic components of the Slurm workload manager used on Stampede2 for batch scheduling of users' compute jobs \cite{slurm}. The three node classes discussed in Section \ref{sysconfig} were configured via Ansible to support the Slurm controller host, Slurm worker hosts and the job submission host on the Jetstream master node VM, compute node VMs, and login node VM, respectively. The Jetstream Slurm controller was configured to tap into a common Slurm database housed within the Stampede2 system. This allowed for inquiries and submission requests to pass from one system to another without the need for any other intermediary service for communication. 

Typically, a user's job on an HPC system will take advantage of at least one of three parallel paradigms including multiprocessing, multithreading, or job packing to utilize as much of the available resources as possible. For workloads that take advantage of internode communication, usually via an MPI resource manager, information needs to be delivered to appropriately allocate tasks across a dynamic set of compute hosts. For an HPC system, this is typically handled internally and automatically via specialized logic in conjunction with information from the workload manager such that a user need not be concerned in setting parameters beyond the number of compute nodes and MPI processes to be used. Thanks to built-in features of the MPI libraries from Intel and MVAPICH, it is possible to bootstrap the Slurm workload manger for its information and launch mechanism such that a user may only need to issue an ``\texttt{mpirun}'' on either Stampede2 or Jetstream systems.

\subsection{Virtual Cluster Creation}

To begin, a persistent master node VM was created to serve as the orchestration point for the rest of the virtual cluster. Other basic configurations included setting up a virtual private network and adding a pair of shared SSH keys automatically upon creation of any new VMs. A non-production node on Stampede2 was designated for serving out the file systems via NFS while the site firewall was configured to allow traffic for this as well as for the Slurm controller to interact with the Stampede2 Slurm database.

Next, a dedicated and persistent login node VM was created as a front end for users to interact with the virtual cluster. With file systems mounted, RPMs installed, and the appropriate Slurm configuration in place, this VM would closely mimic the environment that users would experience on a Stampede2 login node.

From there, compute node VMs ranging from 1 virtual core to 44 could be instantiated via OpenStack commands issued from the master node. As compute nodes were instantiated, each IP was added to an Ansible host inventory to be brought up and ready for use. Updates were first applied, followed by the RPMs in the TACC repository. Next, the Stampede2 file systems were mounted, Slurm was installed and configured, and authorization via LDAP completed the process.

\subsection{Job Submission}

With compute nodes up and running, a user logged into any of the Stampede2 login nodes via SSH could request that a job be directed to either the Jetstream or Stampede2 systems with one additional Slurm command-line option. Similarly, a user logged into the Jetstream login node VM could request that a job either run locally or on Stampede2. A third option, Agave, which provides an API and web portal front end to researchers, was also configured for initial testing and future cloud expansion endeavours.

Concretely, Agave is a set of containerized servers that support a REST API \cite{dooley2012,agave}. It supports computationally intensive research by running and monitoring scientific applications on behalf of a user while recording all inputs, outputs, environment settings, software versions, and hardware used by a job to support experimental \textit{traceability} and \textit{reproducibility}. The main components in Agave are shown in Table \ref{table:components}.

\begin{table}
\centering
\caption{Agave components}
\label{table:components}

\begin{tabular}{ |l|p{5cm}| } \hline
\textbf{Agave Component}  & \textbf{Description}\\ \hline
Execution system & System used for computation where application binaries can be run\\ \hline
Storage system & Data repository that can be accessed through Agave for I/O\\ \hline
Application & Executable code invoked by Agave on a specific execution system\\ \hline
Job & Runtime instance of an application with parameters\\ \hline
\end{tabular}
\end{table}

Existing systems can be defined in Agave as execution or storage systems. Credentials provided to Agave are used to transfer data and run commands on those systems, usually using a protocol like SSH. Web portals such as Designsafe \cite{rathje2017,designSafe} and CyVerse \cite{cyverse} are built on top of Agave to provide researchers a customized graphical user interface to HPC systems. 

%% file: 03-results.tex
\section{Results}
\label{results}

\begin{table}
\centering
\caption{Applications chosen for run time study}
\label{table:apps}

\begin{tabular}{|l|l|p{4cm}|} \hline
\textbf{Name} & \textbf{Version} & \textbf{Description}\\ \hline
GROMACS & 2016.4 & Molecular dynamics simulation for biochemical molecules\\ \hline
NAMD & 2.10 & Molecular dynamics simulation of large biomolecular systems\\ \hline
OpenSeesSP & 2.5.0 & Earthquake simulation and modeling for structural and geotechnical systems\\ \hline
WRF & 3.6.1 & Mesoscale numerical weather prediction system\\ \hline

\end{tabular}
\end{table}

\begin{table*}
\centering
\caption{Run times for applications on HPC and cloud extension systems}
\label{table:result}
\begin{tabular}{|c|c|c|c|c|c|c|c|c|}
\hline
\multirow{ 2}{*}{\textbf{Applications}} & \multicolumn{4}{|c|}{\textbf{Stampede2}} & \multicolumn{4}{|c|}{\textbf{Jetstream}} \\
\cline{2-9}
 & Avg. run time (H:MM:SS) &{CPU} &{Nodes} &{Runs} &{Avg. run time (H:MM:SS)} &{CPU} &{Nodes} &{Runs}\\
\hline
 GROMACS &1:05:40  &8 &4 &3 &1:46:06 &8 &4 &3 \\
\hline
 NAMD\footnotemark[1] &0:02:40 &16 &8 &3 &0:03:58 &16 &8 &3 \\
\hline
 OpenSeesSP\footnotemark[1] &0:03:46 &1 &1 &3 &0:06:43 &1 &1 &3 \\
\hline
 WRF &0:03:50 &4 &2 &2 &0:06:09 &4 &2 &2 \\
\hline
% etc. ...
\end{tabular}
\end{table*}
\footnotetext[1]{Both NAMD and OpenSeesSP were launched directly with Slurm and through Agave's job submission REST API with no difference in run times.}

Several applications were chosen to run on the Jetstream cloud extension system that are regularly executed on Stampede2 and may be flexibly scaled by the number of processors used in computation. Table \ref{table:apps} provides the applications chosen along with version information and a brief description of each. GROMACS \cite{abraham2015}, NAMD \cite{phillips2005}, OpenSeesSP \cite{opensees}, and WRF \cite{skamarock2005} were chosen because input data and historical information were readily available.  All applications were launched via the Slurm \texttt{sbatch} command on both Stampede2 and Jetstream. NAMD and OpenSeesSP were additionally launched with Agave.

To gauge the practicability of seamlessly offloading a typical HPC workload to a cloud system, the same application binaries were run on both Stampede2 and Jetstream. The applications themselves are built as multi-architecture binaries that allow for code branching depending on what level of vectorization instruction is supported on the underlying chip. AVX2 instructions serve as the baseline to support the Haswell architecture on Jetstream while the newer Skylake and KNL architectures of Stampede2 can take advantage of AVX-512 instructions.

The input cases used are described briefly below: For GROMACS, a pure water case was simulated for 1.536 million atoms in the isothermal isobaric (NpT) ensemble at 300K and 1atm with initial coordinates and system parameters coming from the application website for twenty thousand time steps. For NAMD, the APOA1 case is a 92 thousand atom Particle Mesh Ewald (PME) molecular dynamics simulation for three thousand time steps of Apolipoprotein A-1 which is the primary component of the high-density lipoprotein cholesterol molecule.  For OpenSeesSP, the input case conducts transient load analysis of a two-dimensional structure over twenty thousand time steps. For WRF, the input case represents a fixed grid weather forecast at 12km resolution over the continental United States for a 24 hour period.

In this work, we demonstrate the feasibility of HPC workload offload to a cloud extension system. As such, the performance characteristics of interest focus on reasonable relative time to solution when comparing systems. Enough runs to see stable execution times for a handful of canonical input examples were performed on both systems lauching via Slurm and Agave. Results are displayed in Table \ref{table:result}. The Stampede2 runs were conducted on the Skylake architecture and generally outperform runs conducted on Jetstream's Haswell architecture. Eight Jetstream compute node VMs were instantiated with two virtual cores and 4GB of memory each. Compared runs were conducted with the same number of MPI processes per node with no explicit affinity settings.

\begin{table}
    \centering
    \caption{Stampede1 median queue wait times as a percentage of requested run time for varying node counts and requested run times in minutes}
    \label{table:queuetimes}
    \begin{tabular}{|c|c|c|c|c|c|} \hline
\multirow{ 2}{*}{\textbf{Req. Time}}  & \multicolumn{5}{|c|}{\textbf{Requested Node Count}} \\ \cline{2-6} & \textbf{1-4}  & \textbf{4-16}  & \textbf{16-64} & \textbf{64-256} & \textbf{>256}   \\ \hline
1-4       & 3.33\% & 6.67\%  & 8.67\%  & 14.00\%  & 839.67\% \\ \hline
4-16      & 0.00\% & 1.67\%  & 2.00\%  & 14.50\%  & 91.25\%  \\ \hline
16-64     & 0.13\% & 3.67\%  & 1.21\%  & 3.25\%   & 20.13\%  \\ \hline
64-256    & 0.06\% & 9.82\%  & 11.94\% & 25.09\%  & 14.64\%  \\ \hline
256-1024  & 0.34\% & 11.76\% & 6.57\%  & 10.07\%  & 5.59\%   \\ \hline
1024-4096 & 0.67\% & 4.37\%  & 2.91\%  & 3.85\%   & 1.89\%   \\ \hline
    \end{tabular}
\end{table}

Both NAMD and OpenSeesSP were launched directly with Slurm and through Agave's job submission REST API with no difference in run times. Ultimately, Agave launches a job through Slurm's \texttt{sbatch} command after configuring a job's inputs, outputs and other miscellaneous settings. The Agave REST interface hides the details of different HPC schedulers and makes those schedulers accessible from web applications.

%% file: 04-conclusions.tex
\section{Conclusions}
\label{conclusions}

In this paper, we've briefly described the necessary components needed to construct a Jetstream cloud extension that closely emulates a Stampede2 HPC execution environment. Because a portion of the Jetstream and Stampede2 systems are colocated, we were able to exercise administrative privilege to share file systems, network configurations, identity management, and other facility resources that would be more difficult or intractable when interacting in a more general cloud infrastructure. Both systems use the same job scheduler, shell environment, and applications stack that help to provide a consistent user experience. 

With this setup, migrating applications from Stampede2 to Jetstream requires much less work for the user than migrating to an off-site, public cloud. Input, output, application, and library directories are already mounted and ready for use. This also means that VMs are comparatively light weight in their instantiation, minimizing initialization times. File system mounting and network locality further help to avoid potential expensive application and data transfer costs.

The Agave middleware layer provided a high-level interface from which users could launch applications on Stampede2 and on the Jetstream cloud with no additional timing overhead. When interacting with Agave's API, the Jetstream cloud extension is simply another HPC system running Slurm; no additional customization was necessary. This submission technique is an effective mechanism for web portals such as science gateways to dynamically leverage available computing resources.

Four common HPC applications were chosen as part of an initial demonstration for the HPC cloud extension. Multi-architecture application binaries were built to take advantage of the heterogeneous instruction sets of the different systems. Ultimately, time-to-solution was the metric used to establish viability of this experiment. Application runs on the cloud system expectedly result in slower but still acceptable solution times due largely in part to underlying hardware differences. Thus, when HPC queue wait times are long, offloading work to the cloud can both decrease any backlog on the HPC system and can improve end user response time.

\subsection{Future Work}

The results reported in this paper establish the efficacy of on-premises virtualization of HPC systems. Yet, this drive toward automatic cloud bursting raises a number of interesting questions. For instance, more experimentation is needed to determine what policies should be in place to effectively ascertain which applications should even be considered for cloud bursting. Or another, how well would massively parallel computations involving hundreds or even thousands of processors run in this cloud environment? What about offloading jobs involving I/O heavy workloads? Moreover, is there a way to statically qualify or disqualify an application from being considered for cloud execution by inspecting its code, data, and dependencies?

One key factor in deciding whether to move an HPC job to the cloud is whether time-to-solution is improved. If the amount of time between job submission and the availability of the job's results decreases, then, all things being equal, moving to the cloud is beneficial. The time a job waits in the queue can be a substantial percentage of the overall time-to-solution. It's been reported that in one national laboratory, wait times may be up to four times longer than execution times \cite{bicer2011}. Table \ref{table:queuetimes} suggests the median wait times for the retired Stampede1 system tend to be significantly lower than this figure in most cases. While queue wait times follow a heavily skewed distribution towards lower values, there still remains the potential for interacting with the job scheduler and/or historical data to help determine when a job may have a significant wait ahead, providing the conditions in which a cloud burst might be beneficial.

Another consideration involves automation of the cloud bursting process. This initial implementation uses a common Slurm database and command-line flags to transfer jobs from the Stampede2 controller to the Jetstream controller. In the future, it is possible to enable Slurm's federation process that will submit a job to all federated clusters simultaneously only to remove pending duplicates once one of the systems is able to schedule the job. Another possibility would be to construct a job submission filter either via Slurm or Agave to realize a more sophisticated predictive model. One example might be as described by Guo et al. \cite{GuoSSWS14} and would be able to dynamically route jobs to the cloud as HPC backlogs grow.

Future work will include dynamically scaling the number of compute node VMs available based on the HPC system's congestion and the cloud system's current idle resource availability. Finally, adaptations and policy decisions to integrate accounting mechanisms for the two distinct systems will need to be investigated.